\documentclass[12pt]{article}

\usepackage{color}
\usepackage{graphicx}

\usepackage{times}

\title{Long-range incommensurate charge fluctuations in (Y,Nd)Ba$_2$Cu$_3$O$_{6+x}$}

\author{G. Ghiringhelli$^{1, \ast}$, M. Le Tacon$^{2}$, M. Minola$^{1}$,
S. Blanco-Canosa$^{2}$, C. Mazzoli$^{1}$, \\
N.B. Brookes$^{3}$, G.M. De Luca$^{4}$, A. Frano$^{2,5}$, D. G. Hawthorn$^{6}$, F. He$^{7}$, T. Loew$^{2}$, \\
M. Moretti Sala$^{3}$, D.C. Peets$^{2}$, M. Salluzzo$^{4}$, E. Schierle$^{5}$, R. Sutarto$^{7,8}$, \\
G. A. Sawatzky$^{8}$, E. Weschke$^{5}$, B. Keimer$^{2, \ast}$, L. Braicovich$^{1}$\\
\\
\normalsize{$^{1}$ CNR-SPIN, CNISM and Dipartimento di Fisica,}\\
\normalsize{Politecnico di Milano, piazza Leonardo da Vinci 32, I-20133 Milano, Italy}\\
\normalsize{$^{2}$ Max-Planck-Institut~f\"{u}r~Festk\"{o}rperforschung,}\\
\normalsize{Heisenbergstra{\ss}e 1, D-70569 Stuttgart, Germany}\\
\normalsize{$^{3}$ European Synchrotron Radiation Facility, BP 220, F-38043 Grenoble Cedex, France}\\
\normalsize{$^{4}$CNR-SPIN, Complesso MonteSantangelo - Via Cinthia, I-80126 Napoli, Italy}\\
\normalsize{$^{5}$ Helmholtz-Zentrum Berlin f\"{u}r Materialien und Energie,}\\
\normalsize{Albert-Einstein-Stra{\ss}e 15, D-12489 Berlin, Germany}\\
\normalsize{$^{6}$ Department of Physics and Astronomy, University of Waterloo, }\\
\normalsize{Waterloo, Canada N2L 3G1}\\
\normalsize{$^{7}$ Canadian Light Source, University of Saskatchewan,}\\
\normalsize{Saskatoon, Saskatchewan, Canada S7N 0X4}\\
\normalsize{$^{8}$ Department of Physics and Astronomy, University of British Columbia, }\\
\normalsize{Vancouver, Canada V6T 1Z4}\\
\\
\normalsize{$^\ast$To whom correspondence should be addressed;}\\
\normalsize{E-mail: giacomo.ghiringhelli@fisi.polimi.it, b.keimer@fkf.mpg.de.}
}

\date{}

\begin{document}


\baselineskip24pt


\maketitle

\clearpage

\noindent
{\bf There are increasing indications that superconductivity competes with other orders in cuprate superconductors, but obtaining direct evidence with bulk-sensitive probes is challenging.
We have used resonant soft x-ray scattering to identify two-dimensional charge fluctuations with an incommensurate periodicity of $\bf \sim 3.2$ lattice units in the copper-oxide planes of the superconductors (Y,Nd)Ba$_2$Cu$_3$O$_{6+x}$ with hole concentrations $\bf 0.09 \leq$ \emph{\textbf{p}} $\bf \leq 0.13$ per planar Cu ion. The intensity and correlation length of the fluctuation signal increase strongly upon cooling down to the superconducting transition temperature, $\bf T_c$; further cooling below $\bf T_c$ abruptly reverses the divergence of the charge correlations. In combination with prior observations of a large gap in the spin excitation spectrum, these data indicate an incipient charge-density-wave instability that competes with superconductivity.}

A successful theory of high-temperature superconductivity in the copper oxides requires a detailed understanding of the spin, charge, and orbital correlations in the normal state from which superconductivity emerges. In recent years, evidence of ordering phenomena in which these correlations might take on particularly simple forms has emerged \cite{Vojta:AdvPhys2009,He:Science11}. Despite intense efforts, however, only two order parameters other than superconductivity have thus far been unambiguously identified by bulk-sensitive experimental probes: uniform antiferromagnetism in undoped insulating cuprates, and uniaxially modulated antiferromagnetism \cite{Tranquada_Nature1995} combined with charge order \cite{Tranquada_Nature1995, Abbamonte_NatPhys2005} in doped cuprates of the so-called ``214'' family [that is, compounds of composition La$_{2-x-y}$(Sr,Ba)$_x$(Nd,Eu)$_y$CuO$_4$]. The latter is known as ``stripe order'', with a commensurate charge modulation of period $4a$ (where $a = 3.8 - 3.9$ \AA\, is the distance between neighboring Cu atoms in the CuO$_2$ planes) which greatly reduces the superconducting transition temperature, $T_c$, of 214 materials at a doping level $p \sim 1/8$ per planar Cu atom. Incommensurate spin fluctuations in 214 materials with $p \neq 1/8$ \cite{Birgeneau06} have been interpreted as evidence of fluctuating stripes~\cite{Kivelson06}. A long-standing debate has evolved around the questions whether stripe order is a generic feature of the copper oxides, and whether stripe fluctuations are essential for superconductivity.

Recent attention has focused on the ``123'' family [$R$Ba$_2$Cu$_3$O$_{6+x}$ with $R$ = Y or rare earth], which exhibits significantly lower chemical disorder and higher maximal $T_c$ than the 214 system. For underdoped 123 compounds the anomaly in the $T_c$-versus-$p$ relation at $p \simeq 1/8$~\cite{Liang_PRB2006} and the large in-plane anisotropies in the transport properties \cite{Sun_PRL2004,Daou_Nature2010} have been interpreted as evidence of stripe order or fluctuations, in analogy to stripe-ordered 214 materials~\cite{Laliberte_NatCom2011}. Differences in the spin dynamics of the two families have, however, cast some doubt on this interpretation. In particular, neutron scattering studies of moderately doped 123 compounds have revealed a gap of magnitude $\geq 20$ meV in the magnetic excitation spectrum~\cite{Fong2000,Dai2001,Stock2005,Hinkov2011}, whereas 214 compounds with similar hole concentrations exhibit nearly gapless spin excitations~\cite{Birgeneau06}. Further questions have been raised by the recent discovery of small Fermi surface pockets in quantum oscillation experiments on underdoped 123 materials in magnetic fields large enough to weaken or obliterate superconductivity~\cite{Doiron_Nature2007}. Some researchers have attributed this observation to a Fermi surface reconstruction due to magnetic-field-induced stripe order~\cite{Laliberte_NatCom2011}, whereas others have argued that even the high magnetic fields applied in these experiments appear incapable of closing the spin gap, and that a biaxial charge modulation is required to explain the quantum oscillation data~\cite{Harrison_PRL2011}. Nuclear-magnetic-resonance (NMR) experiments have shown evidence of a magnetic-field-induced uniaxial charge modulation~\cite{Wu_Nature2011}, but they do not yield information about electronic fluctuations outside a very narrow energy window of $\sim 1$~$\mu$eV. On the other hand, scattering experiments to determine the periodicity of this modulation and/or to characterize its spin or charge dynamics can currently not be performed under the extreme experimental conditions required to stabilize the high-field phase. The nature of the leading electronic instability of doped 123 materials and its relation to superconductivity and to stripe order therefore remain largely unknown.

We report here the results of resonant x-ray scattering (RXS) experiments especially designed to detect charge order and/or corresponding fluctuations on electronic energy scales. In contrast to prior non-resonant x-ray scattering experiments on the same 123 compounds~\cite{NHAndersen,Strempfer_PRL2004}, the use of photons with energies near the Cu $L_3$ absorption edge ($2p_{3/2} \to 3d$ transitions, $h\nu \simeq 931$~eV) greatly enhances the sensitivity of the scattering signal to the valence electron system \cite{SOM}.
The layered crystal structure of the 123 system (Fig. 1A) and the consequent two-dimensional scattering cross section \cite{SOM} allow for easy variation of the momentum transfer component parallel to the CuO$_2$ planes ($q_{//}$) through rotations of the sample at fixed scattering angle (Fig. 1, B and C). Further experimental parameters include the energy of the incident photons and their polarization
(Fig. 1B). Most of the measurements were carried out with a spectrometer, and the energy transferred by the photon was monitored with a combined instrumental resolution of $\sim 130$ meV. Unlike conventional RXS, we could thus discriminate between, on the one hand, the elastic and quasielastic signal coming from charge density fluctuations, and the dominant inelastic contribution on the other (Fig. 1D) \cite{Ghiringhelli_PRL2004,Braicovich_PRL2010}.

Our central observation is apparent in the raw spectra of underdoped Nd$_{1+x}$Ba$_{2-x}$Cu$_3$O$_7$, shown in Fig. 1D as a function of both photon energy loss and $q_{//}$ along the (1,0) direction, parallel to the Cu-O bonds (Fig. 1B). The inelastic portion of the spectrum comprises features around 100-300 meV resulting from paramagnon excitations \cite{Letacon_NatPhys2011}, and 1-3 eV due to inter-orbital transitions (named $dd$ excitations)
\cite{Ghiringhelli_PRL2004}. These features depend weakly on $q_{//}$, whereas the response centered at zero energy loss exhibits a pronounced maximum at $q_{//} = (0.31, 0)$. Scans around symmetry-equivalent reciprocal-space points \cite{SOM} on multiple samples (see below) have confirmed that the latter feature is generic to the 123 system, and that its intensity maximum is at a wave vector distinctly different from the commensurate position $(1/3, 0)$. We will refer to this feature as a ``resonant elastic x-ray scattering'' (REXS) peak, although, as shown below, it arises from low-energy fluctuations of the valence-electron charge density in the CuO$_2$ planes and is hence not truly elastic.

The 123 structure contains copper atoms in two different crystallographic sites: Cu2 atoms in the CuO$_2$ layers that are common to all high-temperature superconductors, and Cu1 atoms in chains specific to the 123 system that act as charge reservoirs for the layers (Fig. 1A). In x-ray absorption (XAS) and RXS experiments, signals arising from Cu1 and Cu2 sites can be discriminated by virtue of their distinct absorption resonance energy and photon polarization dependence (Fig. 2C) \cite{Hawthorn_PRB2011}. The anomalous REXS peak is present only when the photon energy is tuned to the $3d^9$ configuration of Cu2 sites (Fig. 2, D and E) and can hence be unambiguously assigned to the CuO$_2$ planes. Further confirmation for this assignment comes from a comparison of $q_{//}$-scans along the (0, 1) and (1, 0) directions (parallel and perpendicular to the chains, respectively) in untwinned YBa$_{2}$Cu$_3$O$_{6.6}$ crystals (Fig. 2A). The presence of equally intense REXS peaks in the two directions is incompatible with an origin in the Cu1 chains.

We now use the dependence of the RXS cross section on the polarization geometry to address the question whether the REXS signal arises from a modulation of the charge or spin density in the CuO$_2$ layers. In order to separate these two scattering channels, we measured the scattering intensity for $\sigma$ (Fig. 2A) and $\pi$ polarization of the incident beam, and for the two opposite directions of $q_{//}$ along (1,0) (Fig. 1C). In Figure 2B, we compare the results to model calculations for spin-flip and non-spin-flip RXS from Cu2 sites with $3d^9(x^2-y^2)$ ground state symmetry \cite{Ament_PRL2009,Moretti_NJP2011}. In contrast to the 100-300 meV loss feature, which follows the behavior expected for spin-flip scattering as seen before in undoped and doped cuprates \cite{Braicovich_PRL2010,Letacon_NatPhys2011}, the polarization dependence of the REXS peak indicates that it arises from charge scattering.

Figure 3 provides an overview of the REXS response in a variety of different 123 samples, including both single crystals of YBa$_{2}$Cu$_3$O$_{6+x}$ where the doping level $p$ is adjusted through the density and arrangement of oxygen atoms in the Cu1 chain layer, and thin films of Nd$_{1+x}$Ba$_{2-x}$Cu$_3$O$_7$ where the Cu1 chains are fully oxygenated and $p$ is controlled through the Nd/Ba ratio. The presence of closely similar charge-fluctuation peaks at $|q_{//}| \simeq 0.31$ r.l.u. in both systems in the range $0.09 \leq p \leq 0.13$ confirms that they are independent of the Cu1 chains and other details of the doping mechanism. Samples outside this narrow $p$-range do not show any evidence of the anomalous REXS response, in agreement with prior RXS work \cite{Hawthorn_PRB2011}. This implies weaker charge fluctuations in samples at lower $p$, where incommensurate magnetic order has been observed at low temperatures~\cite{Haug_NJP2010}, and at higher $p$, where the superconducting $T_c$ is maximal. We note that the range in which the REXS peak is observed coincides with the well-known plateau in the $T_c$-versus-$p$ relation of the 123 system (shaded region in Fig. 3), which suggests that competition between superconducting and charge-density correlations is responsible for the anomalously low $T_c$ in this range.

In order to further explore the interplay between superconductivity and the anomalous charge-density response, we have measured the temperature dependence of the REXS peak
both in the energy-resolved mode presented above, and in the more traditional energy-integrated detection (Fig. 4, A and B). The peak is present both above and below the superconducting $T_c$. It narrows continuously upon cooling toward $T_c$, indicating a pronounced increase of the correlation length that reaches $(16 \pm 2) a$ at $T_c$ of YBa$_{2}$Cu$_3$O$_{6.6}$, to be compared with domain sizes in the range $40a$ to $66a$ for stripe-ordered La$_{1.875}$Ba$_{0.125}$CuO$_4$
\cite{Abbamonte_NatPhys2005,Wilkins_PRB2011,Hucker_PRB2011}. The $T$-dependent correlation length demonstrates that the signal arises from charge-density fluctuations, rather than a static charge density wave (CDW). The incipient CDW phase transition heralded by the nearly diverging correlation length is preempted by the superconducting transition, which manifests itself in an abrupt decrease of the intensity and correlation length of the REXS peak (Fig. 4, C and D). This constitutes direct evidence of the competition between superconducting and incommensurate CDW states, which was already suggested by the doping dependence of the REXS response (Fig. 3).
CDW correlations were previously observed by scanning tunneling spectroscopy (STS) on the surface of superconducting Bi$_2$Sr$_2$CuO$_{6+\delta}$ (2201)~\cite{Wise_NaturePhysics2008} and Bi$_2$Sr$_2$CaCu$_2$O$_{8+\delta}$ (2212)~\cite{Hoffman,Howald,Parker_Nature2010},
but only limited information is available from STS on the temperature evolution of the CDW correlations in the 2201 and 2212 compounds, and the relationship of the STS data to the superconducting properties in the bulk of these materials is obscured by electronic inhomogeneity.

The wave vector of the charge correlations revealed in our experiments is in good agreement with the nesting vector of the antibonding Fermi surface sheets predicted by density functional calculations for the 123 system~\cite{Andersen_PhysicaC1991}.
The nesting vector connects those segments on the two-dimensional Fermi surface that develop the maximal gap amplitude in the $d$-wave superconducting state. The CDW is thus a natural consequence of a Fermi-surface instability competing with superconductivity.
The nearly critical nature of the charge fluctuations (Fig. 4)
suggests a CDW ground state for underdoped 123 materials in magnetic fields sufficient to weaken or obliterate superconductivity, which may be responsible for the Fermi surface reconstruction evidenced by the quantum oscillation data~\cite{Doiron_Nature2007}. This scenario agrees qualitatively with a recent NMR study of YBa$_{2}$Cu$_3$O$_{6.5}$ in high magnetic fields, which has revealed signatures of a field-induced CDW \cite{Wu_Nature2011}, and with theoretical work on this issue \cite{Harrison_PRL2011,Yao_PRB2011}. In this precise case, the long range ortho-II oxygen order and/or the high-field conditions may favor the uniaxial commensurate charge modulation with period $4a$ inferred from these data. Further work is required to understand the interplay between the chain order and the in-plane charge modulation,
and to reconcile our results with the NMR data.

It is instructive to compare the charge correlations revealed by our RXS experiments to the $p$-evolution of the spin correlations previously determined by magnetic neutron scattering. For $p \leq 0.08$, these experiments revealed incommensurate magnetic order with wave vectors $q_{//} = (\frac12 + \delta, \frac12)$, where the incommensurability $\delta$ increases monotonically with $p$ \cite{Haug_NJP2010}. Figure 3 shows that neither charge order nor low-energy incommensurate charge fluctuations are present in this doping range. For $p > 0.08$, the spin correlations remain centered at wave vectors similar to those at lower $p$, which bear no simple relation to the wave vector of the REXS peaks determined here \cite{Fong2000,Dai2001,Stock2005,Hinkov2011}. In this doping range, magnetic order disappears \cite{Haug_NJP2010,Coneri,Rigamonti}, and the magnetic excitation spectrum develops a large gap that increases smoothly with $p$, in stark contrast to the abrupt appearance and disappearance of the charge-density correlations with increasing $p$.

These considerations imply that spin and charge order are decoupled in the 123 family, quite different from the 214 family where they coexist microscopically in the ``striped'' state. The small or absent intensity difference between the signal intensities at $q_{//} = (0.31, 0)$ and $(0, 0.31)$ (Fig. 2A) call into question the interpretation of various transport anomalies in the normal state of underdoped 123 compounds in terms of stripe fluctuations \cite{Sun_PRL2004,Daou_Nature2010}, at least in the doping range where we observed the REXS peaks. Although further experiments are required to establish whether these peaks arise from an equal distribution of fluctuating domains of two uniaxial CDWs with mutually perpendicular propagation vectors, or from a single CDW with biaxial charge modulation, the isotropic intensity distribution of the CDW signal shows that these fluctuations cannot account for the strongly anisotropic resistivity and Nernst effect in the normal state.
Despite these materials-specific variations, our data imply that long-range CDW correlations are a common feature of underdoped cuprates. Detailed microscopic calculations are required to assess their relationship to the ``pseudogap'' phenomenon \cite{Vojta:AdvPhys2009}, to the unusual $q=0$ order \cite{Fauqué_PRL06}, and to the polar Kerr effect measurements \cite{Xia_PRL2008} recently reported in some of these materials. The extensive $q$-, $p$-, and $T$-dependent data set we have reported is an excellent basis for theoretical work on these issues.

\bibliographystyle{Science}

\newpage

\begin{figure}
	\begin{center}
		\includegraphics[width=0.9\columnwidth]{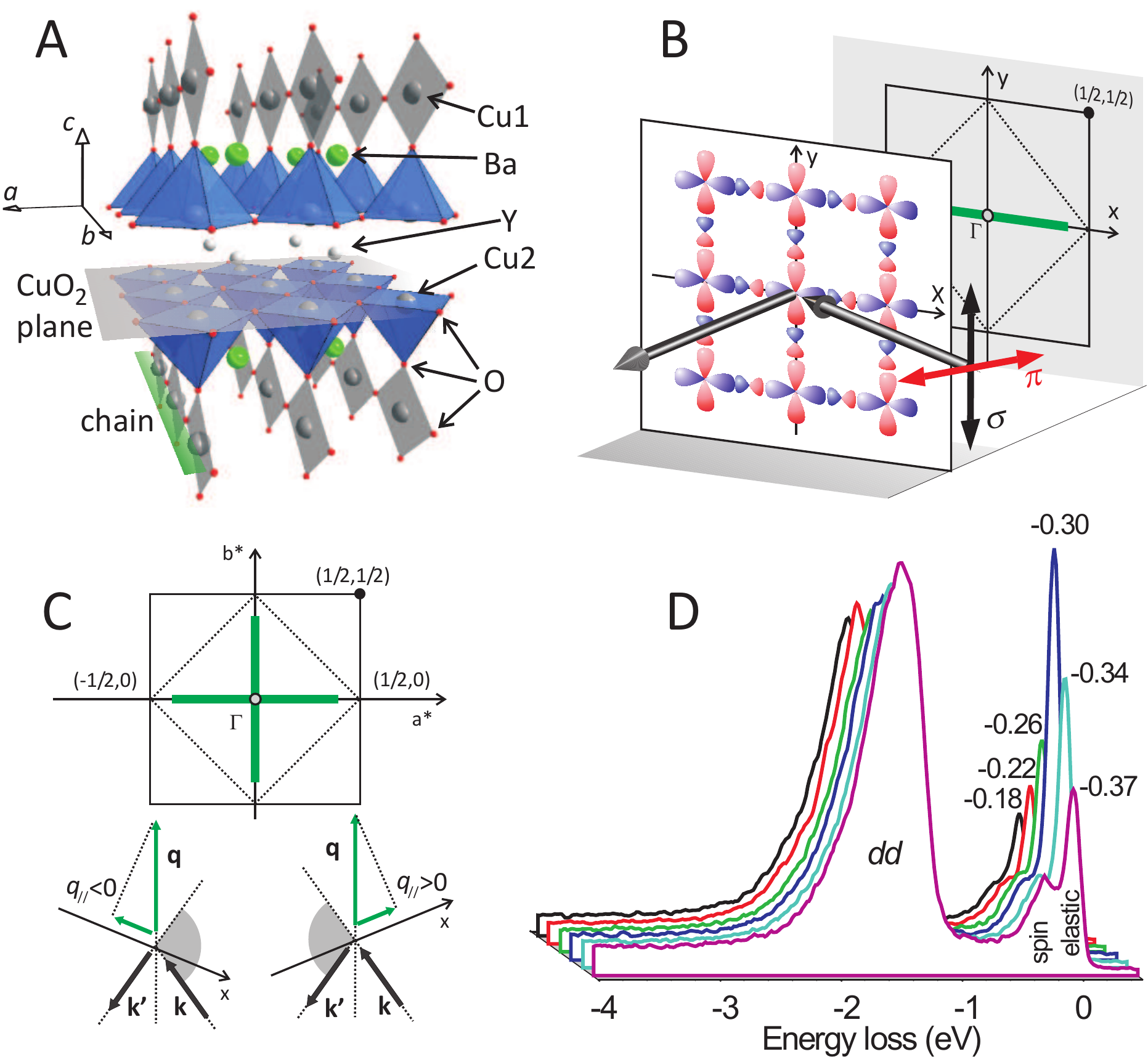}
	\end{center}
	\caption{Resonant soft x-ray scattering from layered cuprates. (A) Crystalline structure of YBa$_{2}$Cu$_3$O$_{6+x}$ superconductors. For $x<1$ some of the O atoms along the chains are missing. (B) Scattering geometry with the $c$ and either $a$ or $b$ axes in the scattering plane. The incident photon polarization can be parallel ($\pi$) or perpendicular ($\sigma$) to the scattering plane. The real and reciprocal spaces are sketched in the front and rear plane respectively. In the real-space image, the Cu $3dx^2-y^2$ and O $2p$ orbitals are shown. In the reciprocal-space image, the nuclear and magnetic first Brillouin zones are drawn with solid and dashed lines, respectively; the thick green line indicates the range covered by the experiments. (C) The in-plane component $q_{//}$ of the transferred momentum ranges from -0.4 to +0.4 reciprocal lattice unit (r.l.u.) along the (1, 0) or (0, 1) directions when the sample is rotated around the $y$ axis at fixed scattering angle, indicated by the grey arcs (130$^\circ$ in most cases). (D) In the Cu $L_3$ energy-resolved RXS spectra of underdoped Nd$_{1.2}$Ba$_{1.8}$Cu$_3$O$_{7}$ ($T_c=65$ K, $T=15$ K, $\sigma$ polarization) the quasi-elastic component has a maximum intensity at $q_{//}=-0.31$ r.l.u.}
	\label{fig:1}
\end{figure}

\begin{figure}
	\begin{center}
		\includegraphics[width=1.2\columnwidth]{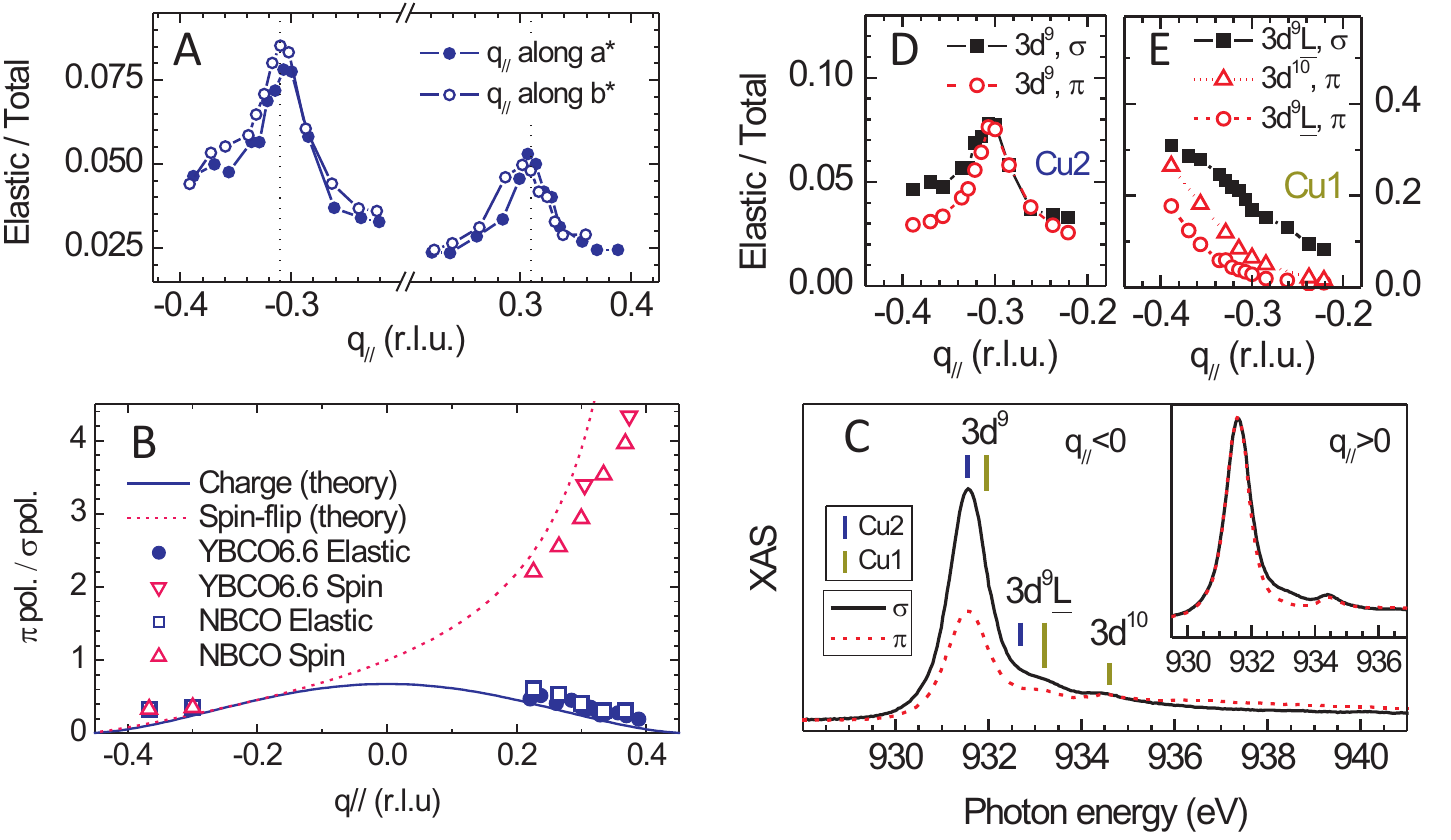}
	\end{center}
	\caption{Polarization and energy dependence of the REXS peak in YBa$_{2}$Cu$_3$O$_{6.6}$. (A) REXS scans measured at the Cu L$_3$ edge for positive and negative values of $q_{//}$ along (1, 0) and (0,1), using $\sigma$ polarization \cite{SOM}.
(B) Ratio between the REXS signal intensities obtained with $\pi$ and $\sigma$ polarizations. The experimental data for YBa$_{2}$Cu$_3$O$_{6.6}$ and Nd$_{1.2}$Ba$_{1.8}$Cu$_3$O$_{7}$ are compared to the model calculations (see text for details) and to the magnetic excitation signal (100-300 meV energy loss). (C) XAS spectra of YBa$_{2}$Cu$_3$O$_{6.6}$ with two polarizations and two geometries corresponding to negative and positive values of $q_{//}$ (Fig. 1C). The main contributions to the XAS peaks are indicated. RXS with photon energies at the main absorption peak of 931.5 eV selects signals arising from the Cu2 sites. (D,E) REXS scans show the CDW peak only when exciting at the Cu2 sites, and nothing at higher excitation energies.
}
	\label{fig:2}
\end{figure}

\begin{figure}
	\begin{center}
		\includegraphics[width=0.9\columnwidth]{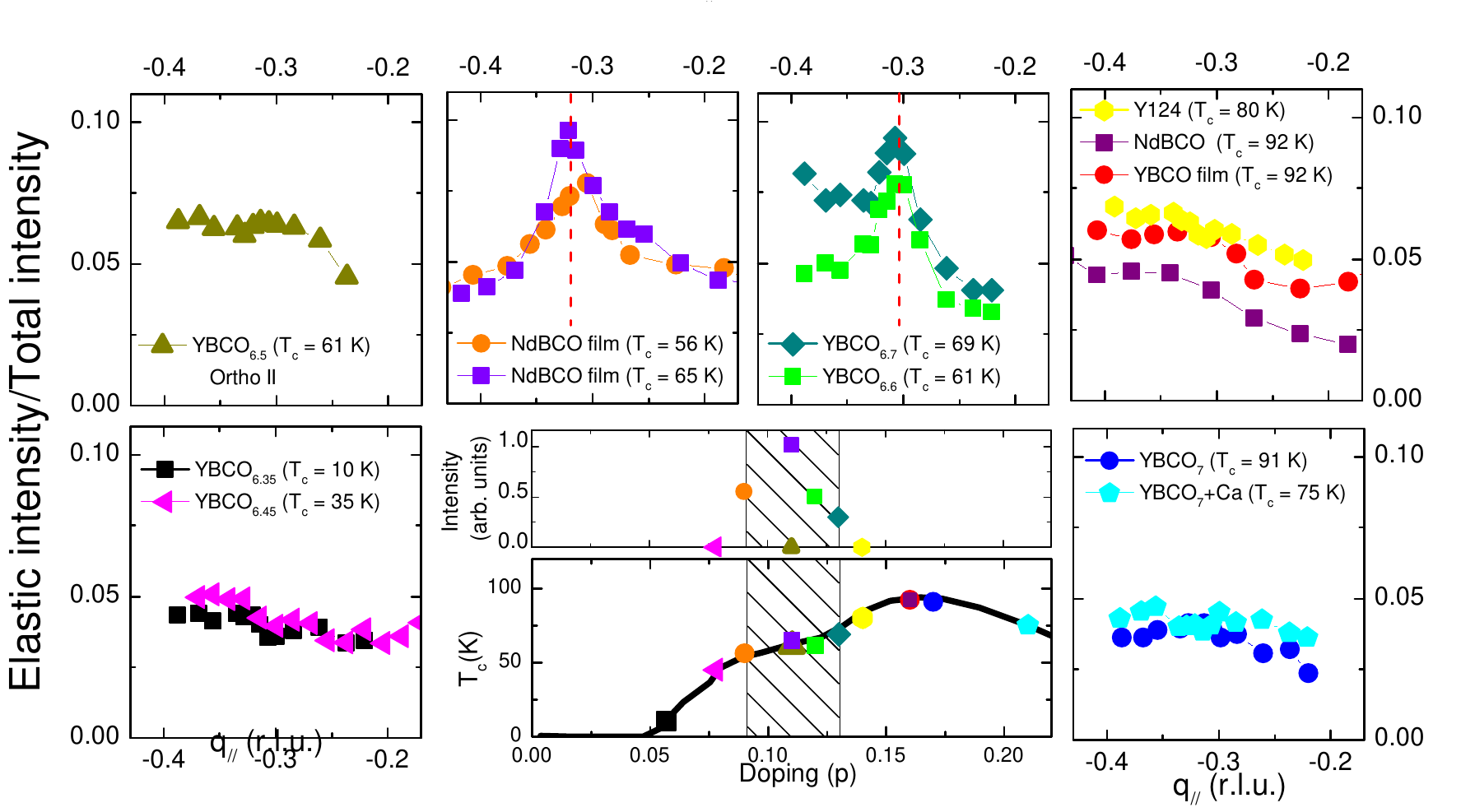}
	\end{center}
	\caption{Dependence of the CDW signal at 15 K on the hole doping level $p$. The CDW signal is present in several YBa$_{2}$Cu$_3$O$_{6+x}$ and Nd$_{1+y}$Ba$_{2-y}$Cu$_3$O$_{7}$ samples, but only for $0.09 \leq p \leq 0.13$. In this doping range (shaded in the central panel), the $T_c$-versus-$p$ relation exhibits a plateau. The CDW peak position does not change with $p$ outside the experimental error, but its intensity is maximum at $p \simeq 0.11$.}
	\label{fig:3}
\end{figure}

\begin{figure}
	\begin{center}
		\includegraphics[width=0.9\columnwidth]{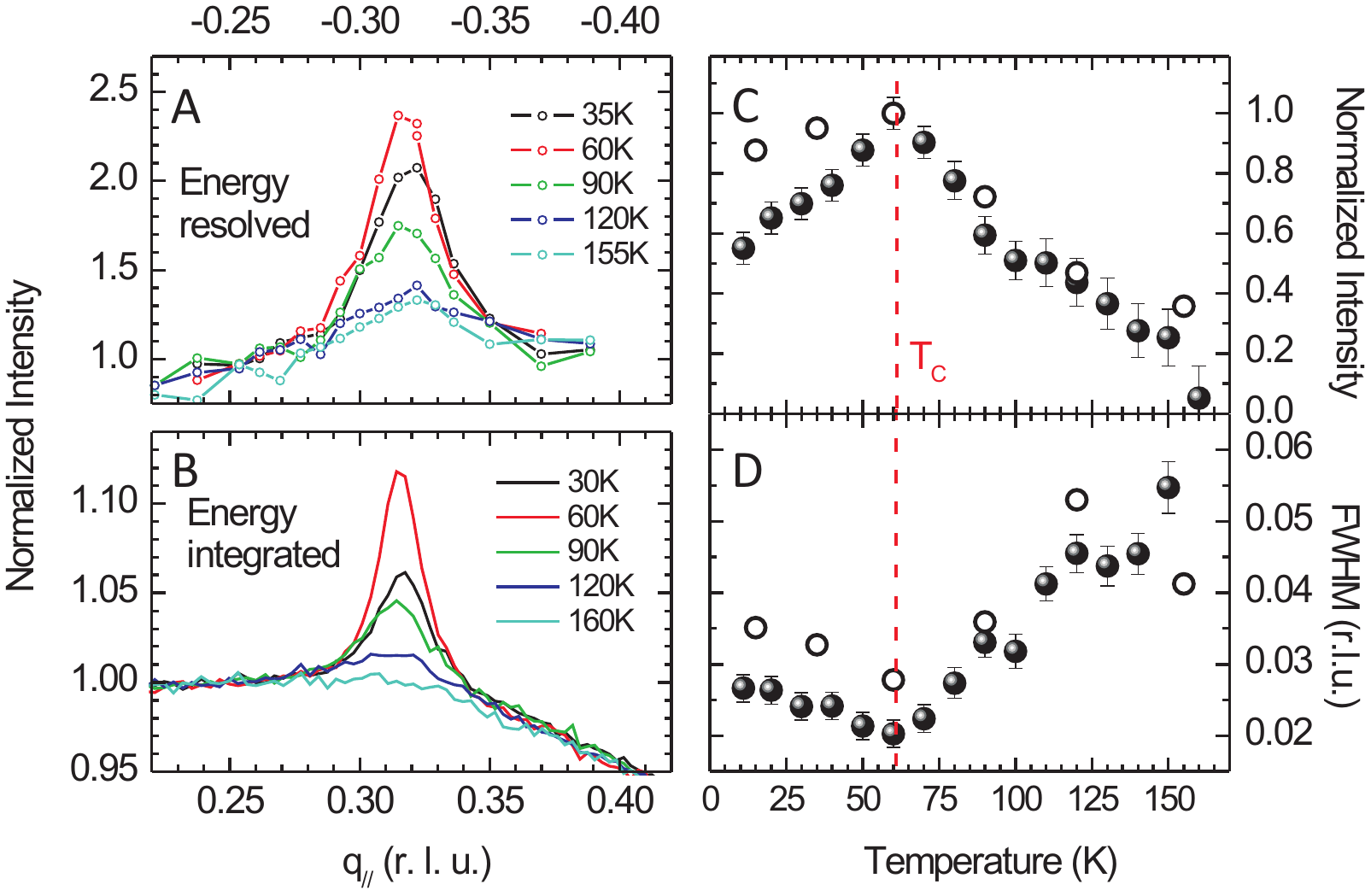}
	\end{center}
	\caption{Temperature dependence of the CDW signal in YBa$_{2}$Cu$_3$O$_{6.6}$. (A,B) Comparison of RXS scans at selected temperatures, as obtained with an energy-resolving instrument and with a conventional diffractometer for soft x-rays. (C,D) $T$-dependence of the CDW intensity and full-width-at-half-maximum (FWHM) derived from the energy-resolved (open circles) and the energy-integrated data (full circles) for $\sigma$ polarization.
}
	\label{fig:4}
\end{figure}

\clearpage

\section*{Supplementary Materials}

\subsection*{Materials and Methods}

\paragraph*{Sample Preparation and Characterization}
We used untwinned YBa$_2$Cu$_3$O$_{6.35}$, YBa$_2$Cu$_3$O$_{6.45}$, YBa$_2$Cu$_3$O$_{6.5}$, YBa$_2$Cu$_3$O$_{6.6}$, YBa$_2$Cu$_3$O$_{7}$ and
Y$_{0.85}$Ca$_{0.15}$Ba$_2$Cu$_3$O$_{7}$ single crystals of volume about $3 \times 3 \times 0.5$ mm$^3$. The corresponding superconducting transition
temperatures $T_c$  were determined for each crystal by SQUID magnetometry (see details in Methods section of Ref.~\cite{Hinkov_Nature04_SOM})
and are listed in Table 1.

The YBa$_2$Cu$_4$O$_8$ single crystals were obtained using KOH flux under ambient pressure in a box furnace. The source material was
conventionally prepared polycrystalline YBa$_2$Cu$_3$O$_7$ mixed with CuO in a molar ratio of 1:1. The details are described in Refs.~\cite{Y124_a, Y124_b}.
The crystal size was $0.8 \times 0.8 \times 0.2$ mm$^3$, and $T_c = 80$ K.
Single crystals were polished in air to obtain a flat and shiny (001) surface.

Nd$_{1.2}$Ba$_{1.8}$CuO$_{7}$ films of thickness 50 nm and 100 nm and  NdBa$_{2}$CuO$_{7}$ and YBa$_{2}$CuO$_{7}$ films of thickness 100 nm were deposited on
SrTiO$_3$ (100) single crystals by diode high-pressure oxygen sputtering. In the former two cases the underdoping is determined by the 1.2/1.8 ratio in the
Nd to Ba relative contents. The two doping levels and $T_c$ are a consequence of the different film thickness, as explained in Ref. \cite{Salluzzo_SOM}. The
optimally doped films were thick enough to neglect the influence of the substrate, which forces the thinner films to grow pseudo-tetragonally. The films had
an area of approximatively 5 $\times$ 5 mm$^2$.
Details of the films growth and characterization can be found in Ref.~\cite{Salluzzo_SOM}. The superconducting transition temperature was determined by DC
four-point transport and by magnetic susceptibility measurements.

The oxygen content and the hole concentration $p$ were determined from the known doping dependence of the out-of-plane lattice parameter $c$ and of $T_c$~\cite{Liang_PRB2006}. For all samples, XAS and RXS measurements were performed without any in-vacuum surface preparation.

\paragraph*{Measurements}
The energy resolved RXS measurements (RIXS) were performed at the ADRESS beam line~\cite{Strocov_JSR2010_SOM} of the Swiss Light Source (Paul Scherrer
Institute, Switzerland) using the \textit{SAXES} spectrometer~\cite{Ghiringhelli_RSI2006_SOM}; and at the ID08 beamline of the European Synchrotron Radiation
Facility (ESRF, Grenoble, France) using the \textit{AXES} spectrometer \cite{AXES,DiNardo}. The XAS measurements were made at the ID08 beamline of the ESRF,
using the fast scanning \textit{Dragon} monochromator. The energy integrated RXS measurements were performed at the UE46-PGM1 beam line of the Bessy-II
storage ring (Helmholtz-Zentrum Berlin, Germany) using the XUV diffractometer.

The resonant conditions were achieved by tuning the energy of the incident x-ray to the maximum of the Cu $L_3$ absorption peak, around 931.5 eV. At that
energy the absorption takes place mainly at the Cu2 sites with $3d^9$ electronic configuration. Although a contribution from Cu1 sites is also present around
the same photon energy as demonstrated in Ref.~\cite{Hawthorn_PRB2011}, we can neglect it because its polarization dependence in the RXS process is expected
to be different from that found in our data (Fig. 2A).
The scattering geometry is shown in Fig. 1 B,C. While keeping the scattering angle ($2 \theta$) fixed, the sample is turned around the $y$ axis,
perpendicular to the scattering plane, thus changing the projection of the transferred momentum $\mathbf{q}=\mathbf{k}-\mathbf{k'}$ onto the direction $x$
parallel to the CuO$_2$ plane. This projection is the meaningful quantity $q_{//}$ for the two-dimensional electron system of the layered cuprates.
At 931.5 eV photon energy the total momentum transfer is $q = 0.944 \sin \theta$ \AA$^{-1}$. In the RIXS experiments $2 \theta = 130 ^\circ$, giving
$q=0.855$ \AA$^{-1}$, which allows one to cover about $\sim 85 \%$ of the first Brillouin zone along the [100] direction (see Fig. 1C).

Momentum transfers are given in reciprocal lattice units (r.l.u.), that is, in units of the reciprocal lattice vectors $a^*$, $b^*$ and $c^*$ where
$a^*$=2$\pi/a$, $b^*$=2$\pi/b$, and $c^*$=2$\pi/c$.
(See table 1 for the values of $a$, $b$, and $c$ for each sample). The conventional sign of $q_{//}$ is shown in Fig. 1C: $q_{//} < 0$ ($q_{//} > 0$) for
grazing incidence onto (emission from) the (001) surface.
The total instrumental energy resolution was about 130 meV, experimentally determined on a non-resonant elastic scattering spectrum of polycrystalline
graphite. The momentum resolution was $\sim 0.005$ r.l.u.  Each spectrum was measured for 5 minutes. The (quasi)elastic intensity was determined by fitting
the zero-energy-loss feature with a Gaussian peak of width not exceeding 1.3 times the instrumental energy width, to take into account possible fluctuations of the energy resolution and uncertainties related count statistics. That intensity was normalized to the
integral of the RIXS spectrum between 0 and 20 eV energy loss (Fig. 1~D). In Fig.~\ref{fig:S2} we have plotted the raw RIXS intensity for the YBa$_{2}$CuO$_{6.6}$ sample at $T \simeq T_c$ = 61 K, where the signal is maximal, for three different momentum transfers: $q_{//}$=0.31 r.l.u. (on the superstructure peak) and  $q_{//}$=0.28 r.l.u. and 0.35 r.l.u. (away from it). The differences between the former and each of the latter spectra are displayed in  Figs.~\ref{fig:S2} B and C. Despite the limited energy resolution, we can safely place an upper bound of 20 meV for the energy of the CDW contribution to the elastic peak.

In the energy integrating experiments $2 \theta = 154 ^\circ$, $q$=0.919 \AA$^{-1}$, $q_{//,\textrm{max}}$=0.44 r.l.u. The scattered intensity was measured
with a photodiode.
In all experimental setups, the alignment of the crystallographic $c$ axis to the scattering plane was made when mounting the sample to the holder, with no
possibility of further adjustment in vacuum. This might lead to inaccuracies when comparing measurements on different samples or along $a^*$ and $b^*$.

\paragraph*{Temperature dependence}
In addition to the data presented in Fig. 4, we have also measured the temperature dependence of the superstructure peak, using the
energy-integrated RXS on the underdoped Nd$_{1.2}$Ba$_{1.8}$CuO$_{7}$ thin film and YBa$_{2}$CuO$_{6.7}$ single crystal. The results are plotted in
Fig.~\ref{fig:S3} and they confirm that the maximum intensity is observed at $T_c$. The systematic determination of the temperature at which where the CDW peak vanishes is complicated by the lesser sensitivity of the energy integrated measurements (weak signal on a large fluorescent background). From the data displayed in Fig.~\ref{fig:S3}, the onset of charge fluctuations appears above $\geq 150$K for the underdoped samples with $T_c$ = 61 and 65 K .

\paragraph*{Additional information}

\begin{figure}[h]
	\begin{center}
		\includegraphics[width=0.5\columnwidth]{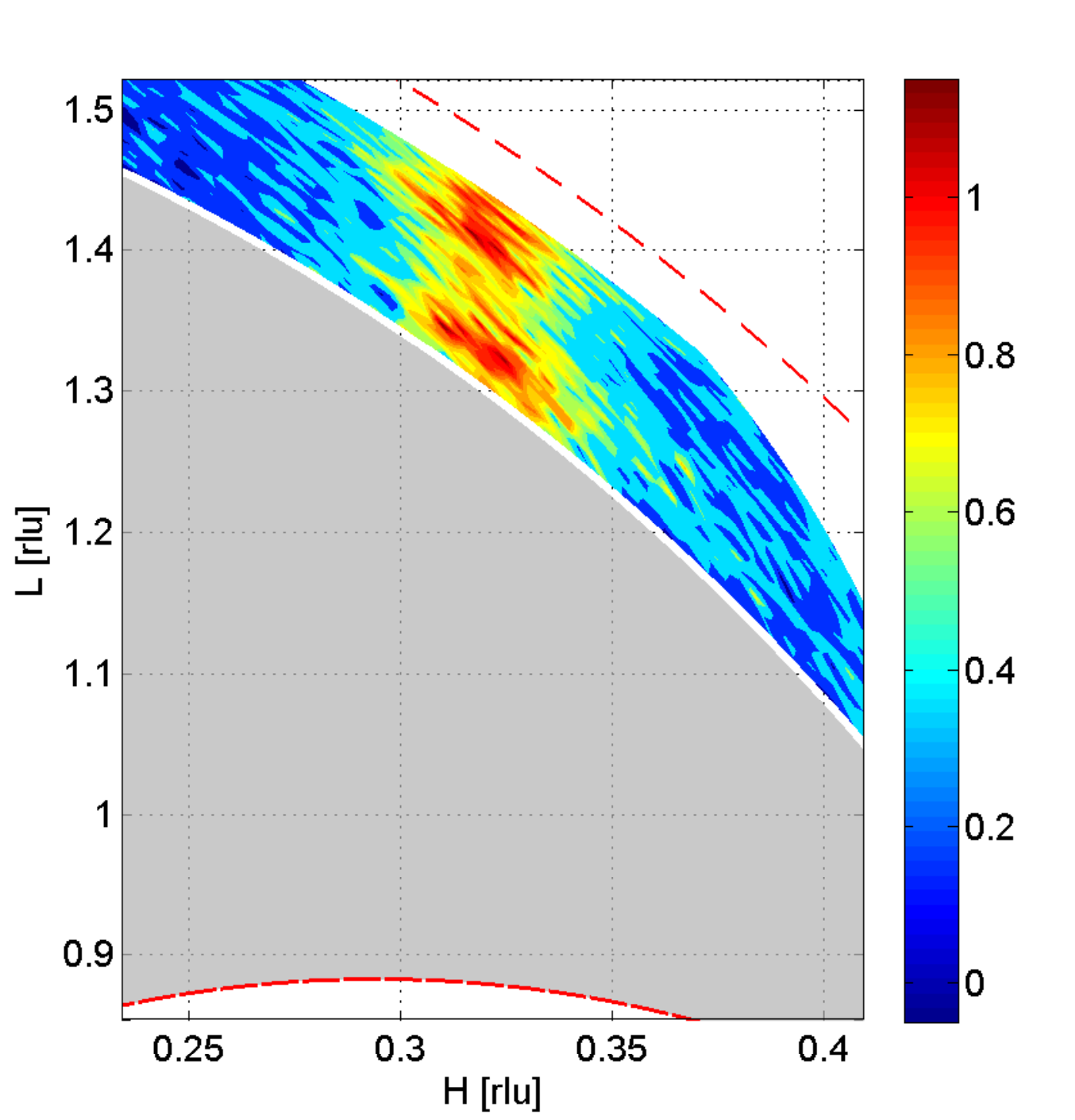}
	\end{center}
	\caption{Color map of the scattering intensity in the $a^* c^*$ plane measured at $T$=15 K in the energy-integrated mode for
Nd$_{1.2}$Ba$_{1.8}$Cu$_3$O$_{7}$, $T_c$ = 65 K. A non constant background, equal to the intensity at 200K,  has been subtracted. The intensity is only
weakly modulated along $c^*$ ($L$ index), and the maximum is at constant $H$ value, indicating that the CDW has intrinsic 2D nature: $H = q_{//}$ is thus the meaningful wave vector in this experiment. The red dashed and dash-dotted line indicate the limits set by the maximum value of $q$ (for $2 \theta = 180 ^\circ$ and the Bragg geometry limit for a (001) oriented surface respectively, for photons of 931 eV energy. Actually the meaningful data collection region is set by the signal to noise ratio, drastically decreasing when moving away from backscattering geometry: the grey area indicates a severe diffuse contamination region for this specific sample, corresponding to $2 \theta _{\textrm{min}} \leq 130 ^\circ$.}
	\label{fig:S1}
\end{figure}

\begin{figure}[h]
	\begin{center}
		\includegraphics[width=0.5\columnwidth]{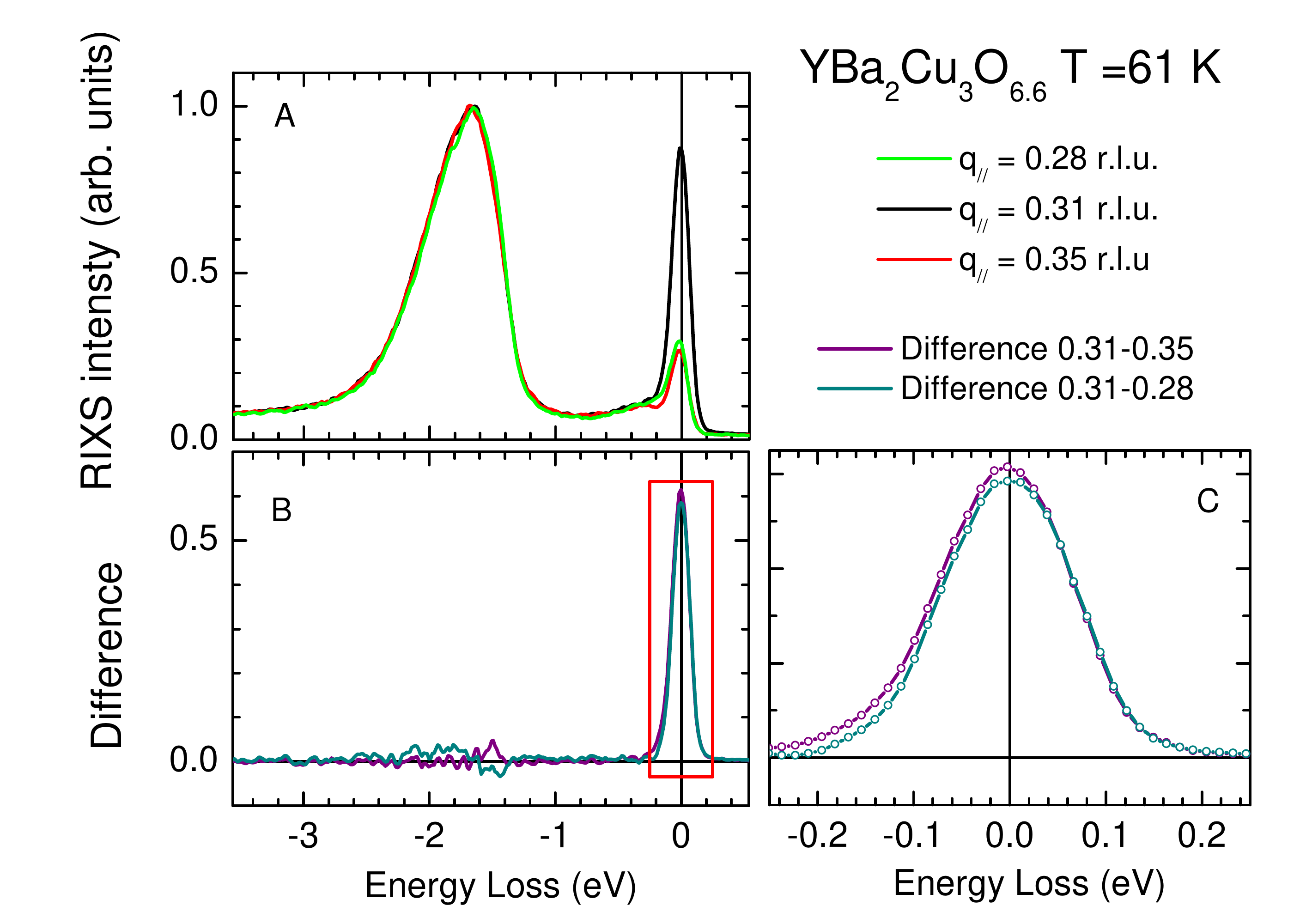}
	\end{center}
	\caption{A) RIXS intensity at $q_{//}$=0.28, 0.31 and 0.35 r.l.u. for the YBa$_2$Cu$_3$O$_{6.6}$ sample at $T \simeq T_c$. B) Difference between the
spectra measured at $q_{//}=0.31$ and 0.28 r.l.u., and $q_{//}=0.31$ and 0.35 r.l.u.. C) Low energy enhancement of the panel B.}
	\label{fig:S2}
\end{figure}

\begin{figure}[h]
	\begin{center}
		\includegraphics[width=0.5\columnwidth]{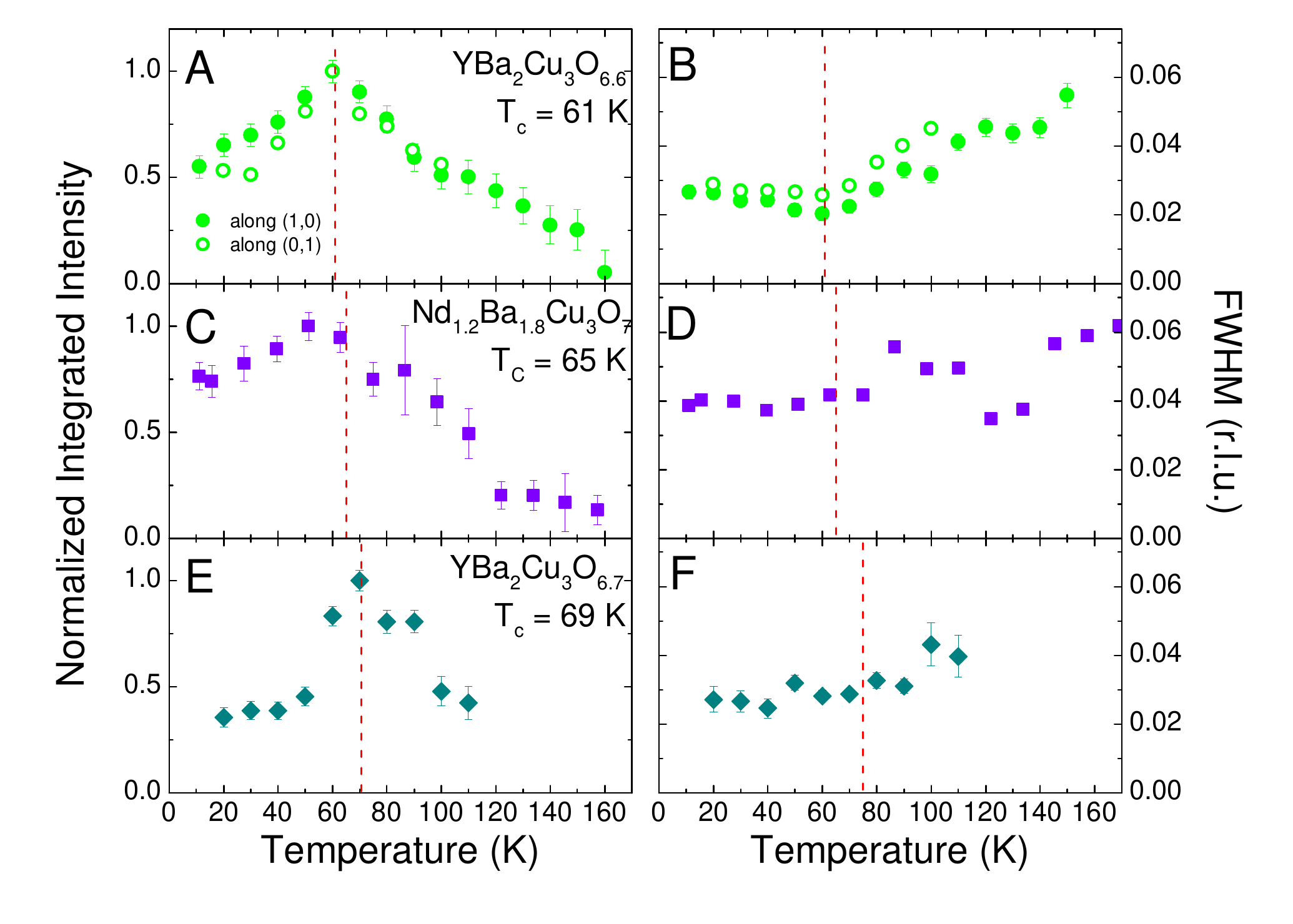}
	\end{center}
	\caption{Temperature dependence of the CDW signal intensity and full-width-at-half-maximum (FWHM) in underdoped YBa$_{2}$Cu$_3$O$_{6.6}$ (A, B), Nd$_{1.2}$Ba$_{1.8}$Cu$_3$O$_{7}$ (C, D), and YBa$_{2}$Cu$_3$O$_{6.7}$ (E, F), measured using energy integrated RXS.}
	\label{fig:S3}
\end{figure}

The two-dimensionality of the CDW was verified in two ways. The comparison of REXS data measured at 130$^\circ$ scattering angle with the diffractometer data
at 154$^\circ$ (Fig. 4~A,B) indicates that the CDW peak is found at exactly the same $q_{//}$ irrespective of the different value of the momentum
perpendicular to the CuO$_2$ planes, (0.31,0,1.29) and (0.31,0,1.44) respectively. We have also measured, for Nd$_{1.2}$Ba$_{1.8}$Cu$_3$O$_{7}$, $T_c = 65
K$, the scattering map in the $(a^*, c^*)$ plane (Fig.~\ref{fig:S1}), which confirms that there is little or no dependence of the scattering cross section along the $c^*$ direction. We can, however, not rule out short-range correlations extending over 1-2 unit cells along the $c$-axis.

\begin{figure}[h]
	\begin{center}
		\includegraphics[width=0.5\columnwidth]{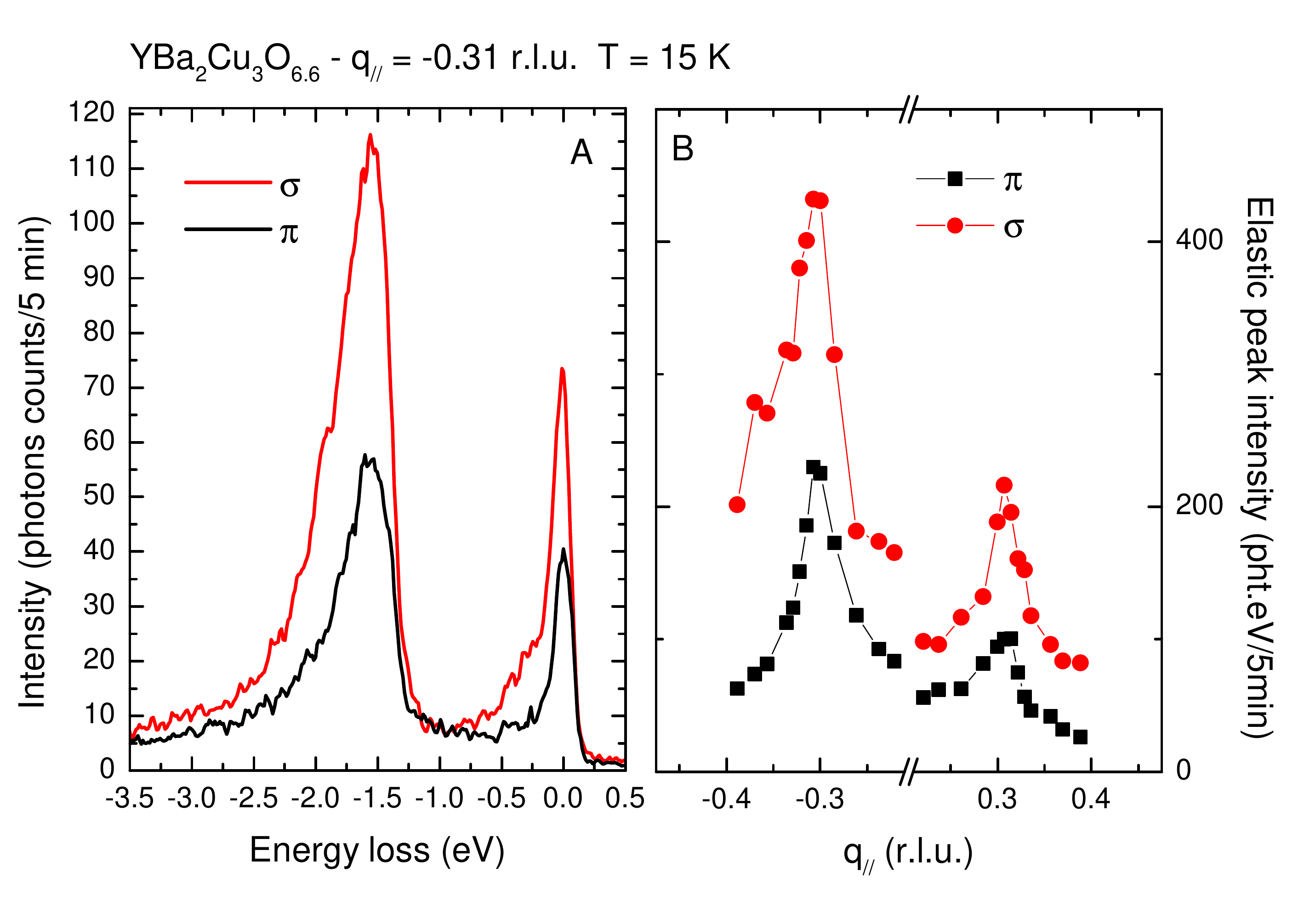}
	\end{center}
	\caption{A) Typical RIXS spectra obtained at the Cu $L3$ edge on YBa$_2$Cu$_3$O$_{6.6}$ in $\sigma$ and $\pi$ polarization for $q_{//}=-0.31$ r.l.u.
		B) Momentum dependence of the integrated intensity of the quasi-elastic line.  The theoretical ratio between the absolute scattering intensities in
the two channels is given by equation~\ref{eq} and displayed in Fig. 2~B, and is in excellent agreement with the experimental points. It must be noted that Fig. 2D has been obtained by normalizing these data to the inelastic intensity, which scales with the absorption cross section similarly to the elastic intensity, thus leading to almost identical curves for $\sigma$ and $\pi$ as shown in Fig. 2D.}
	\label{fig:S4}
\end{figure}

The theoretical lines of Fig. 2~B are obtained from the atomic cross sections for $\pi$ and $\sigma$ incident polarization (illustrated in Fig.~\ref{fig:S4})
as explained in Ref. \cite{Moretti_NJP2011}. Those same cross sections can be multiplied by the structure factor for a 2D antiferromagnetic lattice to obtain
the RIXS intensities from spin waves \cite{Ament_PRL2009} of parent compounds. The atomic cross sections were calculated for a ground state with
$(x^2-y^2)^{\downarrow}$ single hole symmetry and $(x^2-y^2)^{\downarrow}$ and $(x^2-y^2)^{\uparrow}$ final state for the charge and spin-flip scattering
channel respectively. The spin was assumed to be oriented along the (110) direction. In case of spin fluctuations bringing the spin to point in different
directions, the results would not be very different: both channels depend only on the angle $\theta_s$ between the spin and the $c$ axis but not on the
in-plane spin direction. In explicit we have
\begin{equation}
    \frac{I^\mathrm{charge}_{\pi}}{I^\mathrm{charge}_{\sigma}}=\frac{[4 \sin{^2}(\delta - \theta) + \cos ^2 \theta _s] \sin{^2}(\delta + \theta)}{4 + \cos ^2
    \theta _s \sin{^2}(\delta - \theta)}
\end{equation}
\begin{equation}
    \frac{I^\mathrm{spin}_{\pi}}{I^\mathrm{spin}_{\sigma}}=\frac{ \sin{^2}(\delta + \theta)}{\sin{^2}(\delta - \theta)}
\label{eq}
\end{equation}
where $\delta$ is the angle between the $c^*$ axis and $\mathrm{q}$ and $\theta$ is 1/2 of the scattering angle $2 \theta$. In the $\pi / \sigma$ ratios, the
structure factor cancels out \cite{Braicovich_PRB_2010} because it is the same for the two incident polarizations for each final state. In this way we do not
need to make any assumptions about the scattering structure factor when trying to determine whether a given spectral feature originates from spin or charge
scattering, as in Fig. 2~B.
The experimental data in Fig. 2~B were corrected for self-absorption effects. The correction is very small (negligible on the scale of the plot) for $q_{//}
< 0$, and sizable for $q_{//} > 0$. In the latter case $\frac{I^\mathrm{charge}_{\pi}}{I^\mathrm{charge}_{\sigma}}$ was divided by 1.3 - 1.7 and
$\frac{I^\mathrm{spin}_{\pi}}{I^\mathrm{spin}_{\sigma}}$ by 0.55 - 0.75 over the 0.23 r.l.u - 0.37 r.l.u. range. This correction improves the agreement
between theoretical prediction and experimental data, but the very distinct $\pi / \sigma$ intensity ratio of spin and charge channels at positive $q_{//}$
is already clearly visible in the raw data.

We end our discussion with a remark about the experimental methodology of our study. The CDW correlations presented here were first detected on an
energy-resolving RXS spectrometer, where the signal-to-background ratio exceeded 1:1 (Figs. 1~D and 4~A). While this observation was subsequently reproduced on
an energy-integrating RXS diffractometer, the sensitivity obtained there was an order of magnitude lower, due to the large diffuse inelastic background
dominating the RXS signal (Fig. 4~B). The data measured on the former instrument benefit from a 20 times higher signal-to-background ratio, while the energy-integrated scans exhibit a higher overall intensity and hence a lower statistical noise level. This fact can explain why the CDW signal had not been detected before in energy integrated RXS experiments, although made on very similar samples and conditions \cite{Hawthorn_PRB2011}, or with high energy x-ray diffraction, commonly used to study chain ordering and thus also sensitive to CDW \cite{chang}. In addition to its well-known capability to detect spin, charge, and orbital excitations \cite{RMP_RIXS}, our experiments thus establish energy-selective RXS as a highly sensitive probe of momentum-dependent static and quasi-static electronic correlations in transition metal oxides.

\clearpage
\subsection*{Tables}

\begin{table}[h]

\begin{tabular}{|c|c|c|c|c|c|}
\hline
\textbf{Sample}  & $a$ \textbf{(\AA)} & $b$ \textbf{(\AA)} & $c$ \textbf{(\AA)} & $T_c$ \textbf{(K)} & $p$ \\
\hline \hline
YBa$_2$Cu$_3$O$_{6.35}$ &  3.845 &  3.871& 11.781 &  10 & 0.06\\
\hline
YBa$_2$Cu$_3$O$_{6.45}$ &  3.839 &  3.875 & 11.761 &  35  & 0.08\\
\hline
Nd$_{1.2}$Ba$_{1.8}$Cu$_3$O$_7$ (30 nm) & 3.905 & 3.905 & 11.7  & 56  & 0.09 \\
\hline
YBa$_2$Cu$_3$O$_{6.5}$ & 3.829 & 3.875 & 11.731 & 60 & 0.11\\
\hline
Nd$_{1.2}$Ba$_{1.8}$Cu$_3$O$_7$ (100 nm)& 3.885  & 3.92  & 11.7  & 65  & 0.11 \\
\hline
YBa$_2$Cu$_3$O$_{6.6}$ & 3.82 & 3.87 & 11.7  &  61  & 0.12\\
\hline
YBa$_2$Cu$_3$O$_{6.7}$ & 3.826 & 3.880 & 11.709  &  69  & 0.13\\
\hline
YBa$_2$Cu$_4$O$_8$ & 3.84  & 3.87  & 27.25  & 80  & 0.14\\
\hline
YBa$_2$Cu$_3$O$_{7}$ (100 nm) & 3.82 & 3.88 & 11.68 &  91  & 0.16\\
\hline
NdBa$_2$Cu$_3$O$_{7}$ (100 nm) & 3.86 & 3.92 & 11.74 &  92  & 0.16\\
\hline
YBa$_2$Cu$_3$O$_{7}$ & 3.817 & 3.884& 11.681 &  90  & 0.17\\
\hline
Y$_{0.85}$Ca$_{0.15}$Ba$_2$Cu$_3$O$_{7}$ & 3.89 & 3.88 & 11.695 &  75  & 0.21\\
\hline
\end{tabular}

\caption{Composition, lattice parameters ($a$, $b$, $c$), transition temperature ($T_c$) and doping ($p$ holes per Cu2 atom) of the investigated systems.}
\end{table}

\newpage
\clearpage

\subsection*{Figures}

\begin{thebibliography}{99}
\bibitem{SOM} Materials and methods are available as supporting materials on Science Online.
\end{thebibliography}
\end{document}